\documentclass[baaa]{baaa}

\usepackage[pdftex]{hyperref}
\usepackage{subfigure}
\usepackage{natbib}
\usepackage{helvet}
\usepackage[font=small]{caption}

\begin{document}


\journalvol{57}
\journalyear{2014}
\journaleditors{A.C. Rovero, C. Beaugé, L.J. Pellizza \& M. Lares}


\contriblanguage{1}


\contribtype{1}

\thematicarea{9}

\title{Chemical composition of Earth-like planets}


\titlerunning{Chemical composition of Earth-like planets}


\author{M. P. Ronco\inst{1,2}, A. Thiabaud\inst{3}, U. Marboeuf\inst{3}, Y. Alibert\inst{3,4}, G. C. de El\'ia\inst{1,2} \& O. M. Guilera\inst{1,2}}
\authorrunning{Ronco et al.}
\contact{MPR: mpronco@fcaglp.unlp.edu.ar}

\institute{Facultad de Ciencias Astron\'omicas y Geof\'{\i}sicas, Universidad Nacional de La Plata, Argentina \and
  Instituto de Astrof\'{\i}sica La Plata, Consejo Nacional de Investigaciones Cient\'{\i}ficas y T\'ecnicas - Universidad Nacional de La Plata, Argentina \and 
  Physics Institute and Center for Space and Habitability, University of Bern, Switzerland \and
  Observatoire de Besan{\c c}on, France
}


\resumen{
 Los modelos de formaci\'on planetaria se enfocan principalmente en el estudio de los procesos de acreci\'on y 
evoluci\'on din\'amica de los planetas sin tener en cuenta su composici\'on qu\'{\i}mica. En este trabajo, 
calculamos la secuencia de condensaci\'on de los diferentes elementos qu\'{\i}micos para un disco protoplanetario de baja masa alrededor 
de una estrella tipo solar. Incorporamos esta secuencia de elementos qu\'{\i}micos (materiales refractarios y mol\'eculas 
vol\'atiles) en nuestro modelo semianal\'{\i}tico de formaci\'on planetaria que calcula el proceso de formaci\'on de un 
sistema planetario durante su etapa gaseosa. Los resultados de este modelo, distribuci\'on final de embriones y planetesimales, 
son utilizados como condiciones iniciales para desarrollar simulaciones de N cuerpos que computan la formaci\'on post 
olig\'arquica de planetas tipo terrestres. Los resultados de nuestras simulaciones indican que la composici\'on qu\'{\i}mica 
de los planetas que permanecen en la zona de habitabilidad presenta caracter\'{\i}sticas similares a la composici\'on qu\'{\i}mica de la Tierra aunque existen diferencias que pueden deberse al ambiente din\'amico en el cual se formaron.
}

\abstract{
Models of planet formation are mainly focused on the accretion and dynamical processes of the planets, neglecting their
chemical composition. In this work, we calculate the condensation sequence of the different chemical elements for a low-mass 
protoplanetary disk around a solar-type star. We incorporate this sequence of chemical elements (refractory and volatile elements) 
in our semi-analytical model of planet formation which calculates the formation of a planetary system during its gaseous phase. 
The results of the semi-analytical model (final distributions of embryos and planetesimals) are used as initial conditions to 
develope N-body simulations that compute the post-oligarchic formation of terrestrial-type planets. The results of our simulations 
show that the chemical composition of the planets that remain in the habitable zone has similar characteristics to the chemical 
composition of the Earth. However, exist differences that can be associated to the dynamical environment in which they were formed.     
}




\maketitle
\section{Introduction}
\label{S_intro}

\noindent \cite{Alibert2005,Alibert2013} developed a model to calculate the structure and evolution of a protoplanetary disk solving the 
vertical structure of the disk at each radial distance from the central star. Defining the mass of the central star, the mass of 
the disk, and the initial surface density profile, this model allows us to calculate the initial radial profiles of the temperature 
and pressure of the disk. These thermodynamic parameters are then used to calculate the condensation sequence of the 
different chemical elements assuming chemical equilibrium \footnote{To do this we used the comercial software HSC Chemistry for refractory species
and the cooling curve of the disk \citep{Marboeuf2014a} for volatile molecules.}. 
The distribution of chemical elements includes information about the abundance of the most important elements, the formation of 
refractory material and the condensation of volatile molecules, as H$_2$O, CO, CO$_2$, CH$_4$, H$_2$S, N$_2$, NH$_3$ and CH$_3$OH, 
along the disk. All this information is incorporated in our 
semi-analytical model of planet formation \citep{Brunini2008, Guilera2010}, which calculates the formation of a 
planetary system, considering the \emph{in situ} formation of the embryos, during the gaseous phase of the disk. The final 
distributions of embryos and planetesimals calculated by the semi-analytical model are used as initial conditions to carry out N-body 
simulations to compute the post-oligarchic formation of terrestrial-type planets. The results of the N-body simulations allow us to study 
the collisional history of each body in the system and thus, we can determine the final chemical composition of each planet, specially 
those that remain in the habitable zone (HZ).                     
\section{Initial conditions}
Following our previous work \citep{Ronco2014}, we adopted a disk of mass $\textrm{M}_d= 0.03~\textrm{M}_{\odot}$ with an initial gas surface density profile 
given by
\begin{eqnarray}
  \Sigma= \Sigma_0^g \left( \frac{R}{R_c} \right)^{-\gamma} \textrm{e}^{-(R/R_c)^{(2-\gamma)}},
\label{eq1}
\end{eqnarray}
where $R_c= 50$~AU is a characteristic radius of the disk, $\gamma= 1.5$ is the slope of the profile and $\Sigma_0^g$ is a constant of normalization 
which is function of $\gamma$, $R_c$ and the mass of the disk.

\begin{figure}[!ht]
  \centering
  \includegraphics[width=0.34\textwidth]{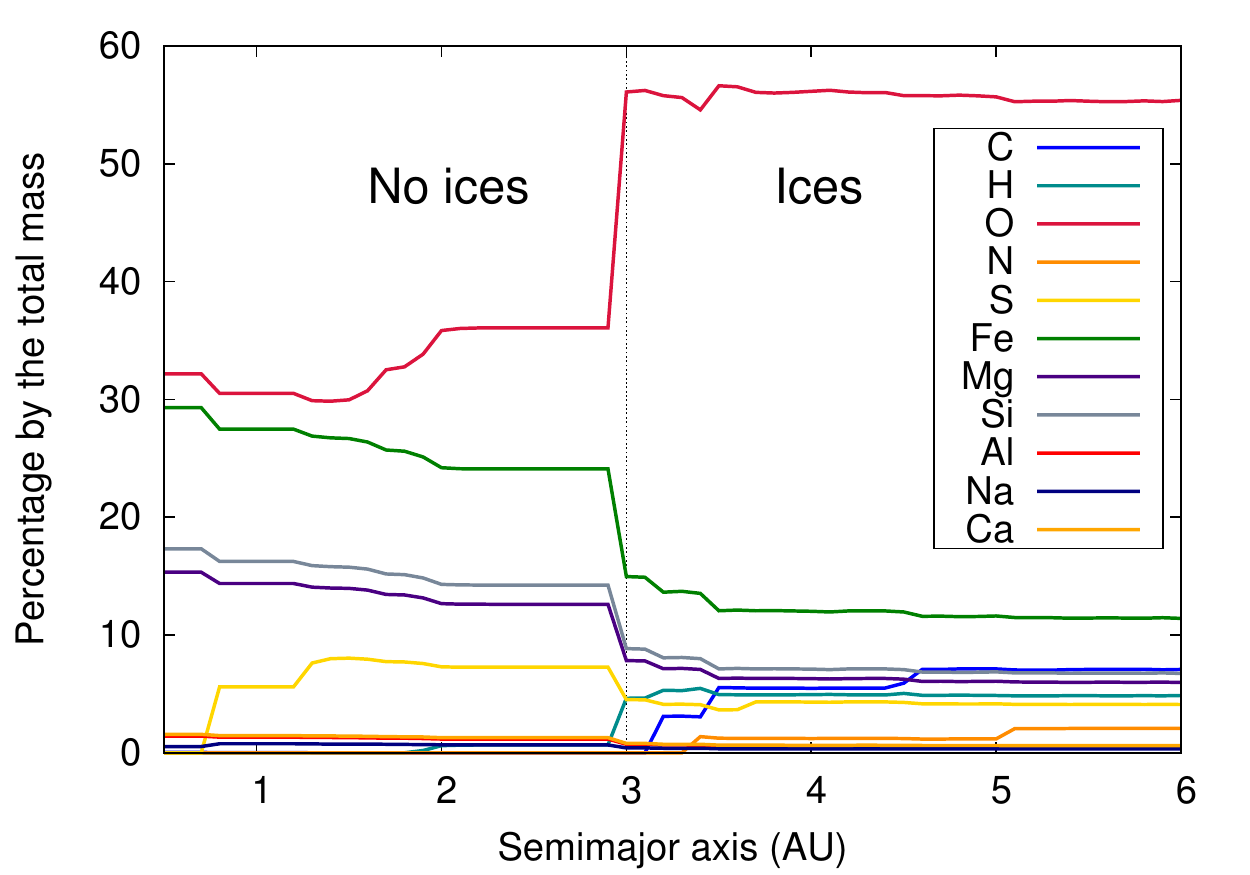}
  \includegraphics[width=0.34\textwidth]{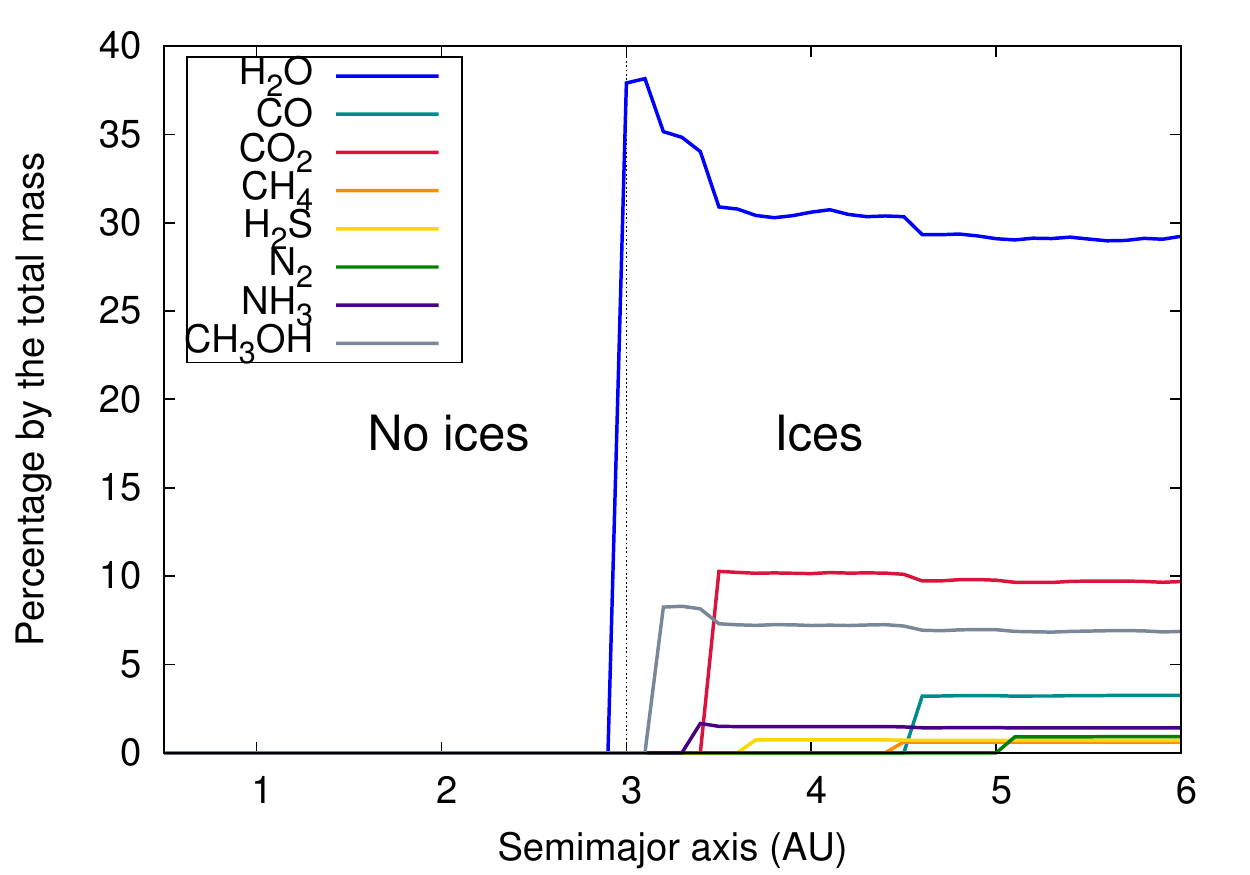}
  \caption{Top panel: Total abundances of the most significant chemical elements that result from the condensation sequence. Bottom panel: Distribution of the most significant volatile molecules (in percentage by mass) product of the condensation sequence. The solid material necessary to complete the total of the mass at each radial distance remains in refractory material. This means that there are not condensed volatile molecules in the inner region delimited by the snow line (located at $\sim 3$~AU and represented by the dashed line).}
  \label{fig:1}
\end{figure}

Using Eq. \ref{eq1}, the model of \citet{Alibert2005} calculated the initial radial profiles of the temperature and 
the pressure. The thermodynamic profile allowed us to calculate the condensation sequences of refractory and volatile 
elements assuming that the disk is initially chemically homogeneous everywhere (see Figure 
\ref{fig:1}). Assuming that the chemical composition of the planetesimals is given by the condensation 
sequences, we thus computed the initial planetesimal surface density. 
It is worth noting that in this work, the snow line is shifted a little bit outside (3~AU according to \citet{Marboeuf2014a}) compared to our 
previous work (2.7~AU according to \citealt{Hayashi1981}). The amount of solid material due to the condensation of water beyond the 
snow line is lower (about 1.75) compared to our previous work (4 following \citealt{Hayashi1981}).

\begin{figure}[!ht]
  \centering
  \includegraphics[width=0.35\textwidth]{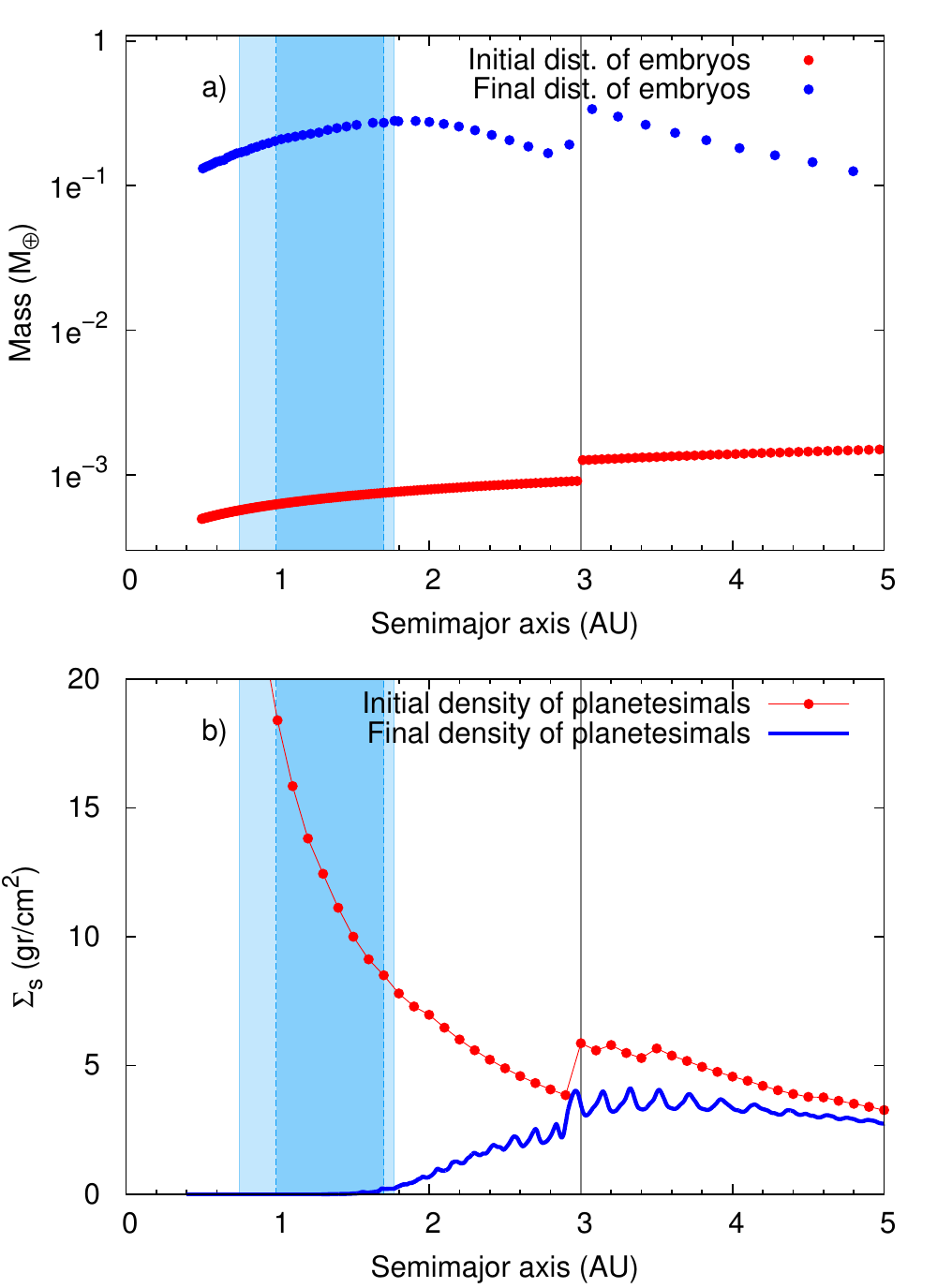}
  \caption{a) Initial (red points) and final (blue points) distributions of embryos calculated with our semi-analytical model. b) Initial (red line-points) and final (blue line) planetesimal surface densities obtained with our semi-analytical model. In both plots, the clear light-blue zone represents the optimistic HZ (between 0.75~AU and 1.77~AU) while the light-blue zone represents the conservative HZ (between 0.99~AU and 1.7~AU) according to \citet{Kopparapu2013}.} 
  \label{fig:3}
\end{figure}

We incorporated these initial surface density profiles (gas and planetesimals) in our semi-analytical model of planet formation and we 
calculated the initial distribution of embryos: the first embryo is located at the inner radius of the disk (0.5~AU) 
and the rest of the embryos are separated by 10 mutual Hill radii (the mutual Hill radius between two bodies of semimajor axis $a_1$ and $a_2$ and
masses $M_1$ and $M_2$ is defined by $R_\text{H,m} = 0.5(a_1 + a_2)[(M_1+M_2)/(3M_\odot)]^{1/3}$) 
until they reached 5~AU; the mass of each embryo is the corresponding 
to the transition mass between runaway and oligarchic growth \citep{IdaMakino1993}. Then, the system evolves until the gas has dissapeared 
(3 Myr). As in \citet{deElia2013} and in a new work in preparation, embryos grow by the accretion of other embryos (when the 
distance between two embryos becomes lower than 3.5 mutual Hill radii we consider perfect merging) and planetesimals. Since embryos are 
formed in situ, their chemical composition will be, for simplicity, the same composition of the accreted planetesimals at a given distance 
from the central star. Therefore, after 3 Myr of evolution, we obtain the distributions of embryos and planetesimals when the gas in the 
disk is dispersed (Figure \ref{fig:3}). These final distributions are considered as initial conditions for the N-body 
simulations to calculate the post-oligarchic growth of the system. From these initial conditions, we generated four simulations distributing 
randomly the orbital elements of the embryos and planetesimals. We used the MERCURY code \citep{Chambers1999} using an integration time-step 
of 6 days in order to calculate with enough precision the most inner orbit. Since terrestrial planets in our solar system might have
formed in 100~Myr - 200~Myr we integrated each simulation for at least 200~Myr.        

\section {Results}
\begin{figure}[!ht]
  \centering
  \includegraphics[width=0.41\textwidth]{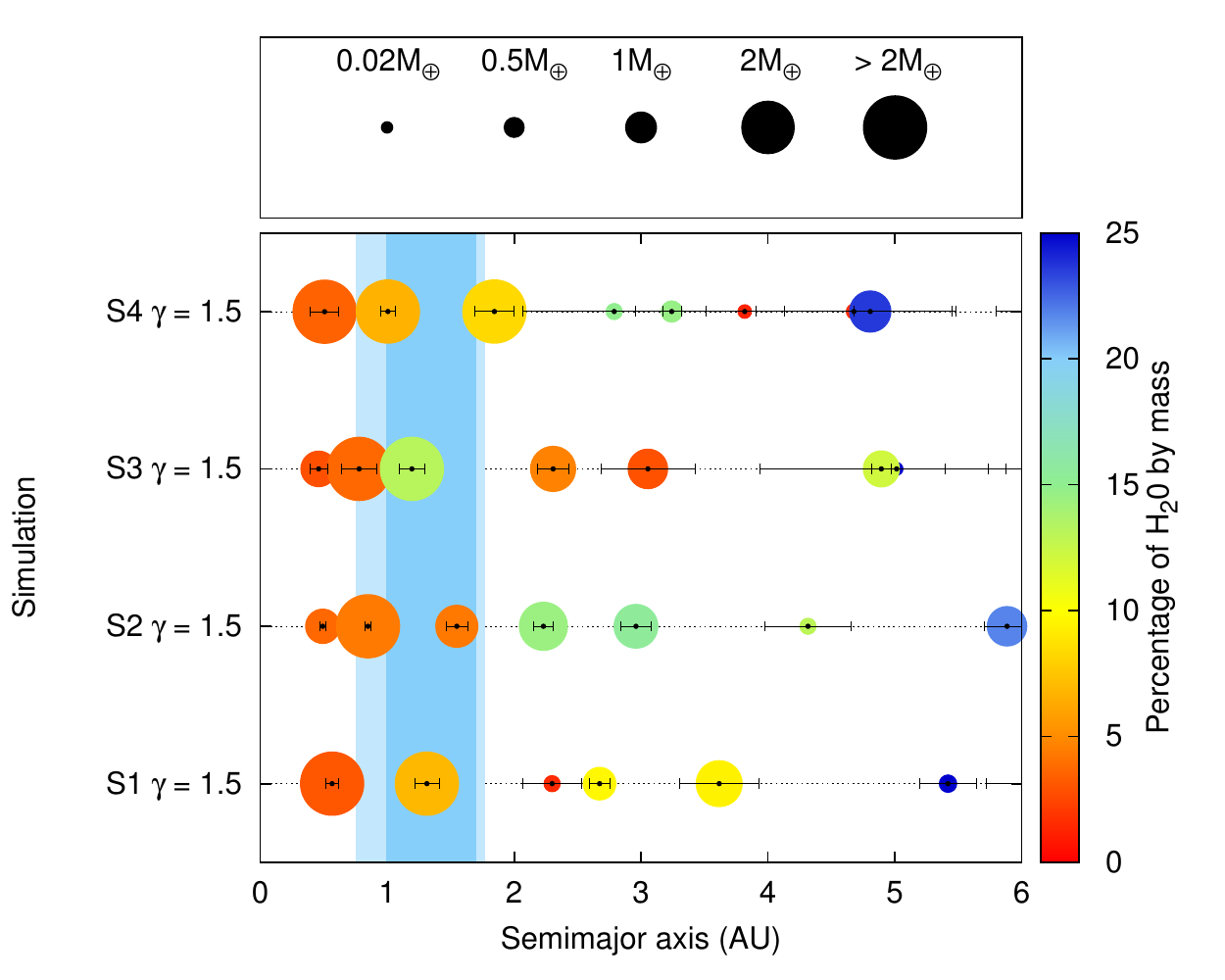}
  \caption{Final configuration of the 4 N-body simulations. The color scale represents the fraction of water of the planets 
    relative to their masses, the shaded regions, the optimistic and conservative HZ. The excentricity of each
    planet is shown over it, by its radial movement over an orbit. All the simulations present planets within the optimistic HZ and their
    water contents range from $4.29\%$ to $13.19\%$ by mass.}
  \label{fig:4}
\end{figure}

All our simulations present planets in the optimistic HZ \citep{Kopparapu2013} with masses ranging from $1.52M_\oplus$ to $4M_\oplus$. The planets
incorporate between $6.37\%$ and $16.41\%$ of volatile material according to the their total mass. Water is the most abundant specie, with values
ranging from $4.29\%$ to $13.19\%$ by mass (Figure \ref{fig:4}).
\begin{table}
\centering
\caption{{Planetary abundances in \% by mass of Mercury and Venus \citep{Morgan1980}, the Earth \citep{KargelLewis1993}, Mars \citep{LoddersFegley1997} 
and the range of values for the planets in the HZ in our simulations.}}
\begin{tabular}{lccccc}
\hline
     & Mercury & Venus & Earth & Mars & Simulations \\
\hline
Fe   &   64.47  & 31.17 & 32.04  & 27.24 &  21.34 - 24.70 \\
O    &   14.44  & 30.9  & 31.67  & 33.75 &  35.23 - 41.93 \\
Mg   &    6.5   & 14.54 & 14.8   & 14.16 &  11.17 - 12.93 \\
Al   &    1.08  &  1.48 &  1.43  &  1.21 &   1.03 - 1.19  \\
Si   &    7.05  & 15.82 & 14.59  & 16.83 &  12.61 - 14.59 \\
Ca   &    1.18  &  1.61 &  1.6   &  1.33 &  0.56 - 0.65   \\
C    &    0.0005&  0.05 & 0.004  &  0.29 &  0.55 - 1.03   \\
Na   &    0.02  &  0.14 & 0.25   &  0.57 &  1.14 - 1.31   \\
\hline
\end{tabular}
\label{tab:1}
\end{table}
In general, the final abundances of the chemical elements obtained in the HZ planets are similar to those 
derived by \citet{KargelLewis1993} (see Table \ref{tab:1}) for the Earth. However, the planets that remain in the optimistic HZ show a mark 
deficit in the final amount of iron as well as an excess of oxygen and carbon.

It is interesting to analyze both, similarities and differences, obtained in the final abundances of chemical elements between the 
resulting HZ planets in our simulations and the Earth. On the one hand, the equivalences obtained may suggest similarities in 
the initial distribution of chemical elements in the protoplanetary disk. On the other hand, the differences could be attributed
primarily to discrepancies associated with the dynamic environment in which planets are formed.
Indeed, unlike what happens with the terrestrial planets in the Solar System, the planets in our simulations
are formed in the absence of gas giants. These differences naturally lead to distinct dynamical histories for terrestrial planets formed 
in both systems, so it is expected to obtain differences in the final abundances of chemical elements. Moreover, the abundances of 
planets formed in our simulations were obtained assuming that collisions are perfectly inelastic. Thus, a more realistic treatment of 
collisions could provide us with a more accurate calculation of the final abundances.

\begin{figure}[!ht]
  \centering
  \includegraphics[width=0.455\textwidth]{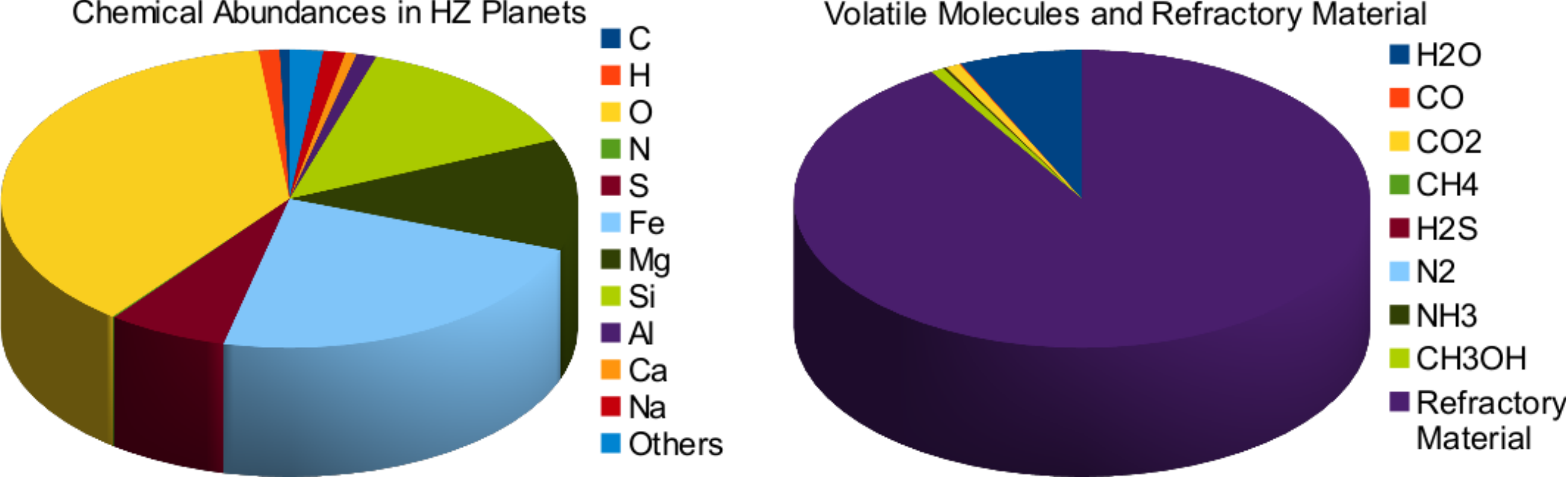}
  \caption{Average chemical abundances and condensed volatile molecules in the final HZ planets.}
  \label{fig:5}
\end{figure}

The abundances of chemical elements of the planets that remain in the optimistic HZ (Figure \ref{fig:5}) are also similar to those
typical abundances found by \citet{Thiabaud2014} for rocky planets around a solar-type star. However, a comparative analysis
between both works should be carried out carefully since the abundances computed for
planets in our simulations are obtained after the dissipation of the gas in the system, and
\citet{Thiabaud2014} analyzed the formation processes of terrestrial planets until the
dissipation of the gaseous component in the disk.

\section {Conclusions}
Starting from initial conditions obtained with a semi-analytical model, we performed
planetary systems with terrestrial planets within the optimistic HZ. After 200 Myr of evolution, the final masses and chemical abundances
of these planets give similar results to that of the Earth. However, exist differences associated to the dinamical environment in which they are formed. In general, the characteristics of the planets that remain in the HZ, particularly their masses and amounts of water,
indicate that they would be potentially habitable planets.

\bibliographystyle{baaa}
\small
\bibliography{Biblio-aaa-Ronco}
 
\end{document}